\documentclass[a4paper]{jpconf}
\usepackage{graphicx}
\begin{document}
\title{Nuclear deformation in the configuration-interaction shell model}

\author{Y. Alhassid,$^{1}$  G.F.~Bertsch$^{2,3}$ C.N.~Gilbreth,$^{2}$ and M.T. Mustonen$^{1}$ }

\address{ $^{1}$Center for Theoretical Physics, Sloane Physics Laboratory,
 Yale University, New Haven, Connecticut  06520, USA\\ $^{2}$Institute of Nuclear Theory,
  Box 351550, University of Washington, Seattle, WA 98915\\
   $^{3}$Department of Physics, Box 351560, University of Washington, Seattle, WA 98195}
\ead{yoram.alhassid@yale.edu}

\def\be{\begin{equation}}
\def\ee{\end{equation}}

\begin{abstract}
 We review a method that we recently introduced to calculate the finite-temperature distribution of the axial quadrupole operator in the laboratory frame using the auxiliary-field Monte Carlo technique in the framework of the configuration-interaction shell model.  We also discuss recent work to determine the probability distribution of the quadrupole shape tensor as a function of intrinsic deformation $\beta,\gamma$ by expanding its logarithm in quadrupole invariants. We demonstrate our method for an isotope chain of samarium nuclei whose ground states describe a crossover from spherical to deformed shapes.  
\end{abstract}

\section{Introduction}
 Deformation is a central concept in understanding the physics of heavy nuclei~\cite{BM75}. However, since intrinsic deformation is introduced by invoking a mean-field approximation that breaks rotational symmetry, it is a challenge to determine the probability density of the intrinsic deformation in the configuration-interaction (CI) shell model, a framework that preserves rotational symmetry. 
 
 Here we review a recent technique we introduced to calculate the axial quadrupole distribution in the laboratory frame using the auxiliary-field Monte Carlo (AFMC) method~\cite{al14,gi17}. We found that this lab-frame distribution exhibits a model-independent signature of deformation. We then discuss recent work in which we used quadrupole invariants~\cite{ku72,cl86} to model the quadrupole shape distribution in the intrinsic frame~\cite{mu17}.   We demonstrate our method for an isotope chain of samarium nuclei, using the model space and interaction of Refs.~\cite{al08,oz13}. Quadrupole invariants were used to extract the effective intrinsic deformation within the framework of the CI shell model in lighter nuclei; see the recent examples in Refs.~\cite{ha16,sc17} and references therein. 

\section{Auxiliary-field Monte Carlo method}
AFMC, also known in the context of the nuclear shell model as the shell model Monte Carlo (SMMC) method~\cite{la93,al94,al17}, 
is based on the Hubbard-Stratonovich (HS) transformation~\cite{hu59}. The Gibbs operator $e^{-\hat H/T}$  of a nucleus described by the Hamiltonian $\hat H$  at temperature $T$  is represented as a superposition of non-interacting propagators $\hat U_\sigma$ of nucleons moving in auxiliary fields $\sigma=\sigma(\tau)$ that depend on imaginary time $\tau$
\be\label{HS}
  e^{-\hat H/T} = \int \mathcal D [\sigma] \; G_\sigma  \hat U_\sigma \;,
\ee
where $G_\sigma$ is a Gaussian weight. The thermal expectation value of an observable $\hat O$ can then be written as
\be\label{observable}
 \langle \hat O \rangle = { {\rm Tr}\,\left(\hat O e^{-\hat H/T}\right) \over  {\rm Tr}\,e^{-\hat H/T}}={{\int {\cal D}[\sigma] G_\sigma \langle \hat O \rangle_\sigma{\rm Tr}\,\hat U_\sigma
\over \int {\cal D}[\sigma] G_\sigma {\rm Tr}\,\hat U_\sigma}} \;,
\ee
 where  $\langle \hat O \rangle_\sigma\equiv {\rm Tr} \,( \hat O \hat U_\sigma)/ {\rm Tr}\,\hat U_\sigma$. The integrands in Eq.~(\ref{observable}) can be calculated using matrix algebra in the single-particle space, and the integration over the large number of auxiliary fields $\sigma(\tau)$ is carried out by Monte Carlo methods. Canonical expectation values at fixed number of protons and neutrons are calculated using a discrete Fourier representation of the particle-number projection~\cite{or94,al99}.  
 
\section{Quadrupole distribution in the laboratory frame}

The lab-frame distribution $P(q)$ of the axial quadrupole $\hat Q_{20}$ at temperature $T$  is defined by
\be\label{Pq} 
P(q )= { {\rm Tr} \left[ \delta(\hat Q_{20} - q) e^{-\hat H/T}\right] / {\rm Tr} \, e^{-\hat H/T}} \;.
\ee
Using compete sets of many-particle eigenstates $| e_m\rangle$ and  $| q_n \rangle$  of $\hat H$ and  $\hat Q_{20}$,  respectively, we have (note that $[H,\hat Q_{20}] \ne 0$)
\begin{equation}\label{prob1}
P(q) = \sum_n \delta(q - q_n) \sum_m \langle q_n |e_m \rangle^2 e^{-e_m/T} \;.
\end{equation}
In the CI shell model, the spectrum of $\hat Q_{20}$ is discrete, but for a heavy nucleus it becomes a quasi-continuum. 

\subsection{Projection on the axial quadrupole}

 To carry out the projection in AFMC, we represent the $\delta$ function as a Fourier integral
 \begin{equation} \label{delta-q}
\delta(\hat Q_{20} - q) = {1 \over 2 \pi} \int_{-\infty}^\infty d \varphi \, e^{-i \varphi q }\, e^{i \varphi \hat Q_{20}} \,,
\end{equation}
and use this in (\ref{Pq}) together with the HS transformation (\ref{HS}) for $e^{-\hat H/T}$.  For each configuration $\sigma$  of the auxiliary fields, we replace the Fourier integral by a discrete Fourier transform. Choosing an interval $[-q_{\rm max}, q_{\rm max}]$, dividing it into $2M+1$ intervals of equal length $\Delta q = 2q_{\rm max} /(2M+1)$, and defining $q_m = m\Delta q$, we have 
\begin{equation}\label{fourier-q}
\Tr\left[\delta(\hat Q_{20} - q_m) \hat U_\sigma \right]  \approx {1\over 2 q_{\rm max}} \sum_{k=-M}^M
  \!\!  e^{-i \varphi_k q_m} \Tr(e^{i \varphi_k \hat Q_{20}} \hat U_\sigma) \;,
\end{equation}
where $\varphi_k = \pi k/q_{\rm max}$ ($k=-M,\ldots, M$). 
Since $\hat Q_{20}$ is a one-body operator, we can calculate the grand-canonical traces on the r.h.s.~of (\ref{fourier-q}) in terms of the matrices ${\bf Q}_{20}$ and ${\bf U}_\sigma$ representing, respectively, $\hat Q_{20}$ and  $\hat U_\sigma$  in the single-particle space, i.e.,  $\Tr\left(e^{i\varphi_k \hat Q_{20}} \hat U_\sigma \right) = 
  \det  \left( 1+   e^{i \varphi_k {\bf Q}_{20}} {\bf U}_\sigma\right)$.

\subsection{Angle averaging}

When using the usual Metropolis algorithm, we find that for a deformed nucleus, the distribution $P(q)$ and its moments are slow to thermalize and have a large decorrelation length.  We resolved this problem by averaging over a specific set of rotation angles $\Omega_j$
\be
\langle e^{i \varphi \hat Q_{20}} \rangle_\sigma \rightarrow {1\over
N_\Omega} \sum_{j=1}^{N_\Omega} \langle e^{i \varphi \hat Q_{20}} \rangle_{\sigma,\Omega_j} \;,
\ee
 where  $\langle e^{i \varphi \hat Q_{20}}  \rangle_{\sigma,\Omega} = {\Tr \left( e^{i \varphi \hat Q_{20}} \, \hat R \hat U_\sigma \hat R^\dagger \right)}/{\Tr \left (\hat R \hat U_\sigma \hat R^\dagger \right )}$, with $\hat R=\hat R(\Omega)$ being the rotation operator with angles $\Omega$.  
We note that any rotation of $\hat Q_{20}$ in (\ref{delta-q}) does not affect the distribution $P(q)$ since the Hamiltonian $\hat H$ is invariant under rotations. The angles $\Omega_j$ are chosen such that $\hat Q_{20}^m$ is proportional to the invariant of order $m$ up to a given order $n$. We have determined a set of 6 angles for $n=2$ and a set of 21 angles for $n=3$~\cite{gi17}. 
All calculations shown here are based on a 21-angle average. We used a time slice of $\Delta \beta=1/64$ MeV$^{-1}$ in a discretized version of the HS transformation (\ref{HS}) and $\sim 5000$ auxiliary-field configurations  for each temperature.

\subsection{Application to samarium isotopes}

 In Fig.~1 we show AFMC distributions $P(q)$ for an isotope chain of samarium nuclei $^{148-154}$Sm at low, intermediate and high temperatures. We observed that the low-temperature distribution for $^{154}$Sm, whose Hartree-Fock-Bogoliubov (HFB) ground state is deformed, is skewed and in qualitative agreement with the distribution for a prolate rigid rotor (dashed line). In contrast, the low-temperature distribution for the spherical nucleus $^{148}$Sm is close to a Gaussian. We conclude that the axial quadrupole distribution in the lab frame is a model-independent signature of deformation.  At low temperatures, we observe a crossover from a spherical to a prolate shape as we increase the number of neutrons. In the isotopes that are deformed at low temperature ($^{150-154}$Sm), we observe a crossover to a spherical shape as we increase $T$. 

\begin{figure}[htb!] \label{Sm_Pq}
  \includegraphics[angle=0,width=0.95\textwidth]{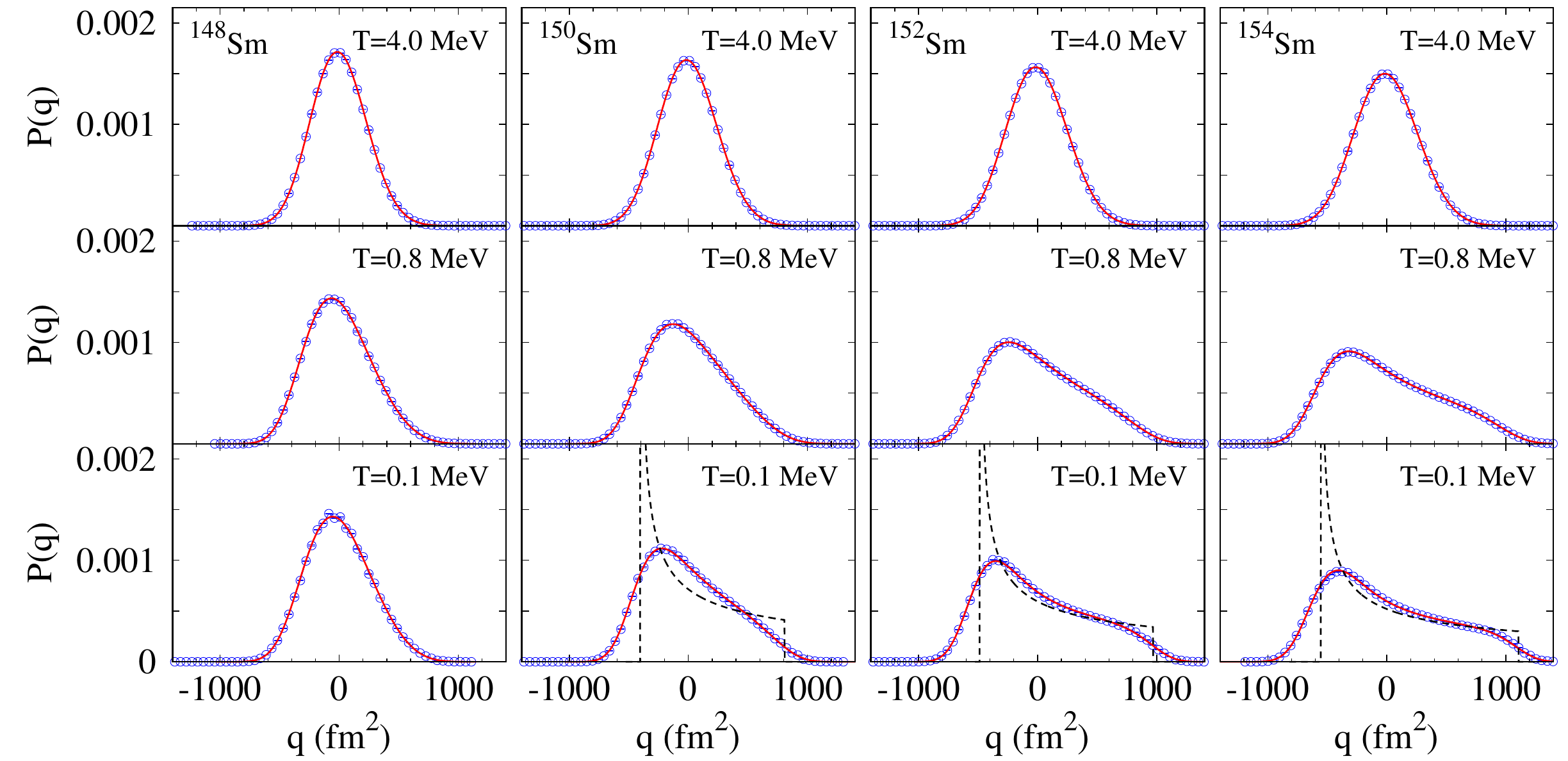}
 \caption{ AFMC distributions $P(q)$ vs.~$q$ (blue circles) for an isotope chain of samarium nuclei at high, intermediate and low temperatures. The dashed lines are rigid-rotor distributions. The red solid lines are the distributions obtained from (\ref{landau}) (see Sec.~\ref{landau_expansion}). Adapted from Ref.~\cite{gi17}.}
\end{figure} 

\section{Quadrupole distribution in the intrinsic frame}

In physical applications, we are interested in the intrinsic deformation of the nucleus.
 Information on intrinsic deformation can be extracted without invoking a mean-field approximation by using quadrupole invariants which are frame-independent. 

\subsection{Quadrupole invariants and their relation to moments of $\hat Q_{20}$.}

A quadrupole invariant is a linear combination of products of the components $\hat Q_{2\mu}$ that is invariant under rotations.  These invariants can be constructed from tensor products of the second-rank quadrupole tensor $\hat Q_{2\mu}$~\cite{ku72,cl86}.  For any given order $2\le n \le 4$, these invariants are unique and their expectation values can be expressed in terms of the corresponding moments of $\hat Q_{20}$:
\be\label{q_invariants}
  \langle \hat{Q} \cdot \hat{Q} \rangle  =  5 \langle \hat{Q}_{20}^2 \rangle \;,\;\;
 \langle (\hat{Q} \times \hat{Q})^{(2)} \cdot \hat{Q} \rangle =   - 5 \sqrt{\frac{7}{2}} \langle \hat{Q}_{20}^3 \rangle \;, \;\;
  \langle (\hat{Q} \cdot \hat{Q})^2 \rangle  =   \frac{35}{3} \langle \hat{Q}_{20}^4 \rangle \;.
\ee
For given values $q_{2\mu}$ of the quadrupole tensor, we define dimensionless quadrupole deformation parameters $\alpha_{2\mu}$ as in the liquid drop model, i.e., $q_{2\mu} = \frac{3}{\sqrt{5 \pi}} 3 r_0^2 A^{5/3} \alpha_{2\mu}$, 
where $r_0 = 1.2$~fm and $A$ is the mass number of the nucleus.  For each set of deformation parameters $\alpha_{2\mu}$, we define an intrinsic frame whose orientation is characterized by Euler angles $\Omega$, and in which the deformation parameters $\tilde \alpha_{2\mu}$ are given by
\be
  \tilde \alpha_{20} =   \beta \cos \gamma, \;\;\;{\tilde \alpha}_{21} = {\tilde \alpha}_{2,-1} = 0, \;\;\; {\tilde \alpha}_{22} = {\tilde \alpha}_{2,-2} = {\rm real}= \frac{1}{\sqrt{2}} \beta \sin \gamma \;.
\ee
The parameters $\beta,\gamma$ are known as the Hill-Wheeler parameters.  The metric of the transformation from the lab-frame variables $\alpha_{2\mu}$ to the intrinsic-frame variables $\beta,\gamma,\Omega$ is given by
 \be\label{metric}
\prod_\mu d {\alpha_{2\mu}}=  \frac{1}{2}\beta^4 |\sin (3\gamma)|  \, d \beta \, d \gamma \, d \Omega \;.
\ee
Quadrupole invariants can also be constructed from $\alpha_{2\mu}$, and up to fourth order, they are given by 
\be\label{alpha_invariants}
  \alpha \cdot \alpha =  \beta^2 \;,\;\;\; [\alpha  \times \alpha ]_2 \cdot \alpha  = -\sqrt{\frac{2}{7}} \beta^3 \cos (3\gamma) \;,\;\;\;
  (\alpha  \cdot \alpha )^2 = \beta^4 \;.
\ee

\subsection{Landau-like expansion}\label{landau_expansion}

The distribution $P(T,\alpha_{2\mu})$ of the quadrupole deformation $\alpha_{2\mu}$ at temperature $T$ is a rotational invariant and therefore it depends only on the intrinsic parameters $\beta,\gamma$.  Using a Landau-like expansion~\cite{al86}, we expand the logarithm of $P$ in the quadrupole invariants up to fourth order 
\be\label{landau}
  P(T, \beta, \gamma) = \mathcal{N}(T) e^{ -a(T) \beta^2 - b(T) \beta^3 \cos (3\gamma) - c(T) \beta^4 }\;,
\ee
where $a,b,c$ are temperature-dependent coefficients and $\mathcal{N}$ is a normalization constant  
determined  from $4\pi^2  \int  \,d {\beta} \,d {\gamma} \, \beta^4 |\sin (3\gamma)| P(T,\beta,\gamma) =1$. 
The parameters $a,b,c$ are determined by matching the expectation values [calculated with the distribution (\ref{landau})] of the three quadrupole invariants as expressed in~(\ref{alpha_invariants}) with their AFMC values, which can be computed from the corresponding moments of $P(q)$ using Eqs.~(\ref{q_invariants}). 

\subsection{Validation of the Landau-like expansion}

To test the validity of (\ref{landau}), we construct the lab-frame distribution $P(T,\alpha_{2\mu})$ by expressing the quadrupole invariants in terms of the lab-frame deformation $\alpha_{2\mu}$ [see Eq.~(\ref{alpha_invariants})]. We then integrate over all $\alpha_{2\mu}$ with $\mu \ne 0$ to find the lab-frame distribution of $\alpha_{20}$, or equivalently $P(q)$, and compare it with the AFMC distribution. The distributions $P(q)$ calculated from the model~(\ref{landau}) are shown by the solid red lines in Fig.~1 and are in excellent agreement with the AFMC distributions (open blue circles). 

\subsection{Application to samarium isotopes}

\begin{figure}[bth]
\includegraphics[width=\textwidth]{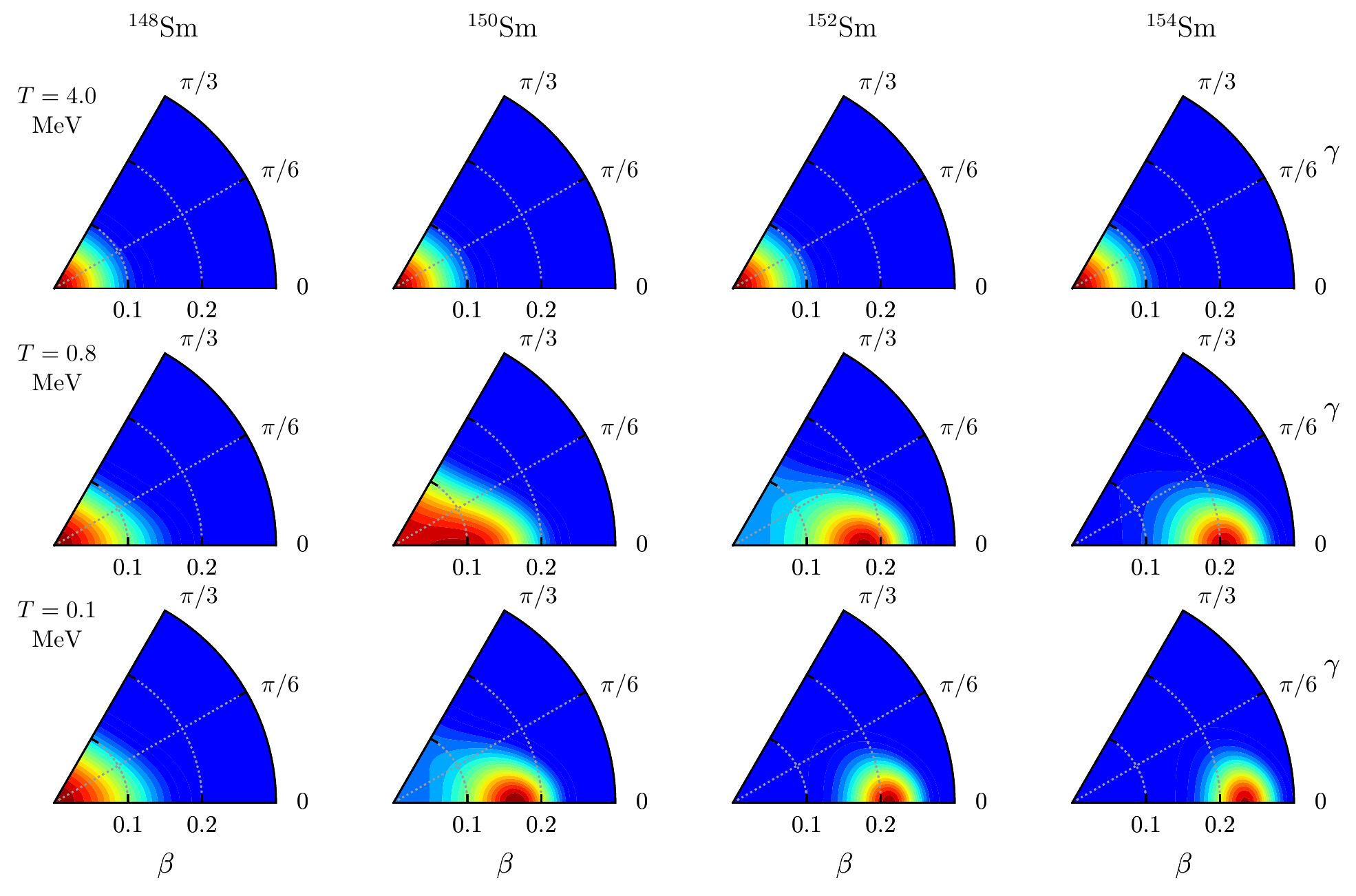}\hspace{2pc}%
\caption{\label{probability_distribution} Intrinsic shape distributions $P(T,\beta,\gamma)$ at low, intermediate and high temperatures for the even-mass samarium isotopes $^{148-154}$Sm. Adapted from Ref.~\cite{mu17}.}
\end{figure}

Figure \ref{probability_distribution} shows the calculated shape distributions $P(T,\beta,\gamma)$ defined in~(\ref{landau}) vs.~$\beta,\gamma$ for the samarium isotopes at the same temperatures as in Fig.~1. The maxima of the distributions (\ref{landau}) mimic the shape transition observed in the HFB mean-field approximation but in the framework of the CI shell model~\cite{note}. As a function of neutron number we observe a transition from a spherical to prolate shape, while nuclei that are deformed in their ground state make a transition from deformed to spherical shape as a function of temperature. 

\begin{figure}[bth]
\begin{minipage}{14pc}
 \includegraphics[width=0.65\textwidth]{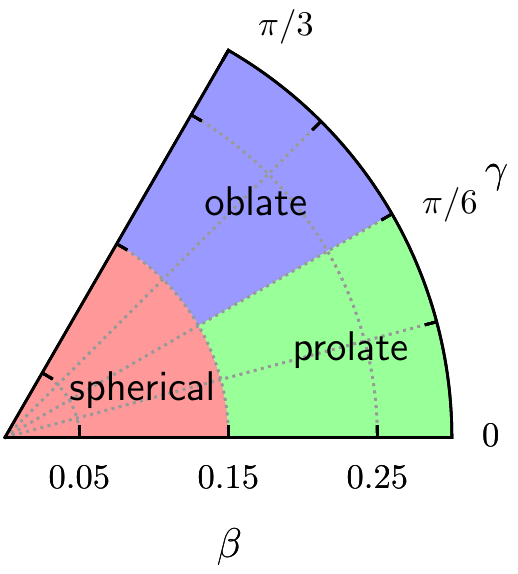}\hspace{2pc}%
 \end{minipage}
\begin{minipage}[b]{20pc}\caption{\label{illustration_shapes} Definition of a spherical, prolate and oblate shape regions in the $\beta-\gamma$ plane. These regions are used in presenting the results of Fig.~\ref{pshapes}. Taken from Ref.~\cite{mu17}.}
\end{minipage}
\end{figure}

\begin{figure}[bth]
\includegraphics[width=\textwidth]{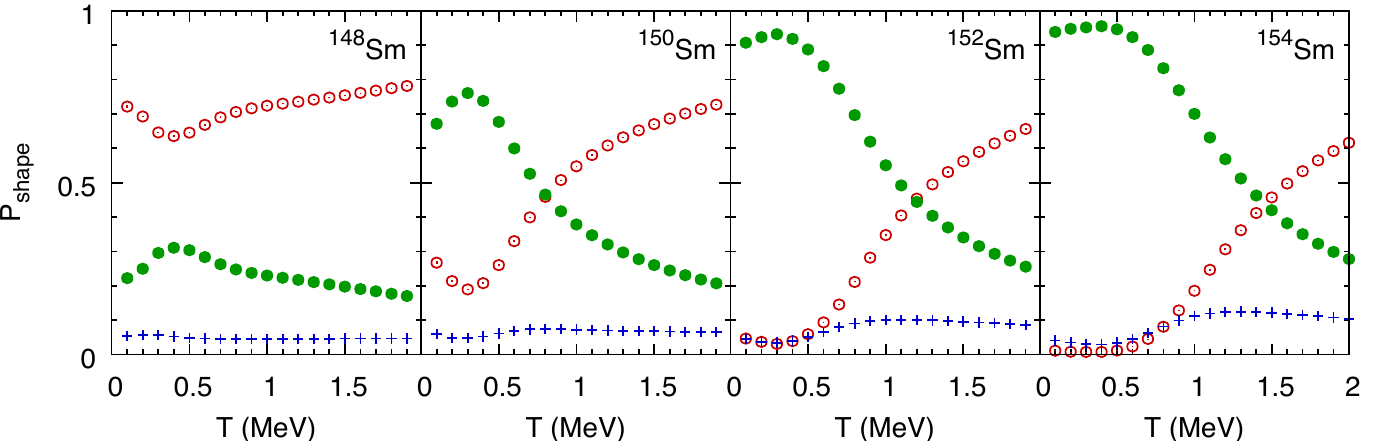}
\caption{\label{pshapes} Probabilities of spherical (red open circles), prolate (green solid circles), and oblate (blue pluses)  regions as a function of $T$  for $^{148-154}$Sm isotopes. Adapted from Ref.~\cite{mu17}.}
\end{figure}

To simplify the presentation of our results we divide the $\beta$-$\gamma$ plane into the three regions as shown in Fig.~\ref{illustration_shapes}, which we choose to represent spherical, prolate and oblate shapes.  For each region, we then define $P_{\rm shape}(T)$ to be the probability to find the nucleus in the corresponding region, i.e., $P_{\rm shape}(T) =  4\pi^2  \int_{\rm shape}  \,d\beta\,d\gamma \, \beta^4|\sin 3\gamma| P(T,\beta,\gamma)$. In Fig.~\ref{pshapes} we show these probabilities as a function of temperature $T$ for the four even-mass samarium isotopes. In the spherical $^{148}$Sm, the spherical region dominates at all temperatures, while in the deformed $^{152,154}$Sm isotopes, the prolate region has a probability close to $1$ at low temperatures and the spherical region becomes the most probable above a certain temperature. The transitional nucleus$^{150}$Sm exhibits an intermediate behavior.

\section{Conclusion and outlook}
We discussed a method we recently introduced to calculate lab-frame and intrinsic shape distributions within the CI shell model without invoking a mean-field approximation. Using the saddle-point approximation, it is also possible to convert the finite-temperature intrinsic shape distribution (\ref{landau}) to  level densities $\rho(E_x,\beta,\gamma)$ as a function of excitation energy $E_x$ and intrinsic deformation $\beta,\gamma$~\cite{mu17}.  Deformation-dependent level densities are useful in the modeling of nuclear shape dynamics, such as fission. 

\section*{Acknowledgments} 
This work was supported in part by the U.S. DOE grant Nos.~DE-FG02-91ER40608 and DE-FG02-00ER41132.
The research presented here used resources of the National Energy Research Scientific Computing Center, which is supported by the Office of Science of the U.S. Department of Energy under Contract No.~DE-AC02-05CH11231.  This work was also supported by the HPC facilities operated by, and the staff of, the Yale Center for Research Computing.


\begin{thebibliography}{}
\bibitem{note} Note, however, that the maxima of the probability density distribution in $\beta,\gamma$, given by $4\pi^2 \beta^4|\sin 3\gamma| P(T,\beta,\gamma)$, do not exhibit sharp shape transitions.
\end{thebibliography}
\end{document}